# Quantum Long Short-Term Memory for Drug Discovery

Liang Zhang[2]†, Yin Xu[1,2]†, Mohan Wu[1], Liang Wang[1] and Hua Xu[1]*

† denotes co-first authors

[1]College of Artificial Intelligence, Tianjin University of Science and Technology, Tianjin 300457, China

[2]Yiwei Quantum Technology Co., Ltd, Hefei 230088, China

Corresponding Author:

H. Xu

Email: hua.xu@ywquantum.com

**Abstract**

Quantum computing combined with machine learning (ML) is a highly promising research area, with numerous studies demonstrating that quantum machine learning (QML) is expected to solve scientific problems more effectively than classical ML. In this work, we present Quantum Long Short-Term Memory (QLSTM), a QML architecture, and demonstrate its effectiveness in drug discovery. We evaluate QLSTM on five benchmark datasets (BBBP, BACE, SIDER, BCAP37, T-47D), and observe consistent performance gains over classical LSTM, with ROC-AUC improvements ranging from 3% to over 6%. Furthermore, QLSTM exhibits improved predictive accuracy as the number of qubits increases, and faster convergence than classical LSTM under the same training conditions. Notably, QLSTM maintains strong robustness against quantum computer noise, outperforming noise-free classical LSTM in certain settings. These findings highlight the potential of QLSTM as a scalable and noise-resilient model for scientific applications, particularly as quantum hardware continues to advance in qubit capacity and fidelity.

## Introduction

ML has achieved remarkable success across various scientific disciplines, including material science[1, 2, 3], computational chemistry[4], and drug discovery[5, 6]. In particular, deep learning (DL) has emerged as a prominent research focus over the past decade, driving numerous breakthroughs and gaining widespread adoption in industry. Its impact extends across both technological advancements and societal development. Notably, AlphaFold, developed by Google DeepMind, has demonstrated highly accurate protein structure predictions, representing a major breakthrough in computational biology[7].

Drug discovery is a costly, complex, and time-consuming process that plays a vital role in human health and well-being[8, 9]. While experimental exploration remains the most reliable approach for identifying novel compounds with key properties, numerous ML and DL methods have been proposed to enhance efficiency and success rates. However, significant improvements are still needed for practical applications[10]. A major bottleneck lies in their computational demands—most DL models rely on high-dimensional data and complex representations[11], which require substantial computational resources and prolonged training times[12]. As models grow in size and complexity, their computational needs escalate exponentially, yet available resources remain limited. This constraint significantly slows the pace of drug discovery, hindering the rapid identification of new molecules and materials with potential scientific and medical breakthroughs.

Quantum computing has demonstrated exponential speedups in certain computational tasks, showcasing its potential for processing ultra-large-scale data[13]. Recent studies have highlighted the growing role of quantum technologies in scientific applications, including full quantum tomography studies on Google's Sycamore processor[37] and the exploration of Noisy Intermediate-Scale Quantum (NISQ) devices[38]. Moreover, recent reviews emphasize that quantum computing offers promising capabilities for efficiently navigating vast chemical spaces and accurately modeling molecular interactions—two essential aspects of early-stage drug screening and lead optimization[32, 36]. The ability of quantum computing to handle high-dimensional and complex data representations presents significant advantages, positioning it as a key direction in the evolution of ML and DL. Hence, the application of quantum computing in scientific fields, particularly drug discovery, is expected to bring about substantial and transformative changes.

Various quantum machine learning (QML) models have been proposed for drug discovery tasks, including quantum neural networks (QNN), quantum support vector machines (QSVM), and quantum generative adversarial networks (QGAN). While QGANs are particularly effective in generating novel molecular structures[33, 34], they are primarily designed for generative tasks, not classification. In contrast, QNNs and QSVMs are focused on classification and regression problems, leveraging quantum kernels or parameterized quantum circuits (PQC) to process high-dimensional data[35].

Long short-term memory (LSTM)[14], a type of recurrent neural network (RNN),

has made significant contributions to drug discovery[5, 6]. Recently, Chen *et al*. proposed a quantum version of LSTM, the quantum long short-term memory (QLSTM) model, which demonstrated improved performance and faster convergence compared to its classical counterpart[15]. Sipio *et al.* also successfully utilized the QLSTM model to perform the parts-of-speech tagging task via numerical simulations, achieving the same overall performance as the purely classical model while using only half the parameters[16]. Notably, although LSTM was originally designed for sequential data, here QLSTM functions as a nonlinear feature transformation layer for static molecular fingerprints. Its gated architecture captures complex feature interactions and dynamically modulates information flow, even without temporal dependencies. This makes QLSTM particularly suitable for high-dimensional, structured data such as ECFP fingerprints.

Building on this approach, we have demonstrated that the QLSTM is capable of handling molecular screening tasks, showing higher prediction accuracy and faster convergence compared to the classical LSTM across multiple datasets. We also observed that the performance of the QLSTM improves as the number of qubits increases. Additionally, we have introduced noise into the QLSTM model to evaluate its effectiveness on currently available NISQ devices. Our investigations revealed that the typical noise of current NISQ devices does not significantly affect the overall model performance.

**Method**

We conducted systematic experiments using five benchmark datasets relevant to drug discovery tasks. Three of these—BACE, Blood-Brain Barrier Penetration (BBBP), and Side Effect Resource (SIDER)—were sourced from MoleculeNet[17], a widely recognized benchmarking suite for molecular machine learning. These datasets address key challenges in early-phase drug development[48, 49], including target-specific activity prediction (BACE), pharmacokinetic property estimation (BBBP), and adverse effect profiling (SIDER). In particular, BACE includes 1,522 molecules labeled for β-secretase 1 inhibition, a therapeutic target for Alzheimer's disease. BBBP contains 2,053 compounds annotated for their ability to penetrate the blood-brain barrier, a crucial factor for central nervous system drugs. SIDER comprises 1,427 approved drugs linked with adverse drug reactions, supporting toxicity risk assessment across multiple organ systems. To further evaluate model performance on diverse data sources, we also incorporated two breast cancer datasets derived from in vitro drug screening: BCAP37 and T-47D. These datasets represent distinct molecular subtypes—triple-negative breast cancer (TNBC) and Luminal A, respectively—and contain 275 and 313 samples. All molecular structures were represented using SMILES strings, which were then encoded into ECFP via the RDKit toolkit[19, 20], with a radius of 6 and fingerprint length of 1024 bits. This configuration follows best practices established in previous benchmark studies, such as MoleculeNet[17], and is also implemented in widely-used deep learning frameworks like Chemprop[50], ensuring compatibility and comparability across molecular representation pipelines.

LSTM is a classic ML model which has been widely applied across various domains and industries due to its ability to effectively handle sequential data. The QLSTM model, a quantum counterpart of LSTM, replaces the classical neural networks in the LSTM cells with a Variable Quantum Circuit (VQC) (Figure 1). The VQC consists of three main components: data encoding, a variational layer, and quantum measurement. The data encoding circuit maps classical vectors into quantum states using angle encoding, where each feature is mapped to the rotation angle of an RX gate. This approach results in shallow circuits and enhanced noise robustness, making it well-suited for NISQ devices. Alternative encoding schemes, such as amplitude encoding, are promising for future exploration but currently face challenges due to increased circuit depth and complex state preparation. The variational layer, consisting of trainable circuit parameters, serves as the model's learnable component and is updated via gradient descent algorithms (Figure 2). Finally, quantum measurements extract output values for subsequent processing. The mathematical equations of the QLSTM model are defined as:

$$f_t = Sigmoid(VQC_1(v_t)) \tag{1}$$

$$i_t = Sigmoid(VQC_2(v_t)) \tag{2}$$

$$\tilde{C}_t = Tanh(VQC_3(v_t)) \tag{3}$$

$$c_t = f_t * c_{t-1} + i_t * \tilde{C}_t \tag{4}$$

$$o_t = Sigmoid(VQC_4(v_t)) \tag{5}$$

$$h_t = VQC_4(o_t * tanh\,(c_t)) \tag{6}$$

where $Sigmoid$ and $Tanh$ are the activation functions, $f_t$ is the forget gate, $i_t$ is the input gate, $o_t$ is the output gate, $v_t$ is a concatenation of the input at step t and the hidden state at step t-1 and $h_t$ is the hidden state of QLSTM model.

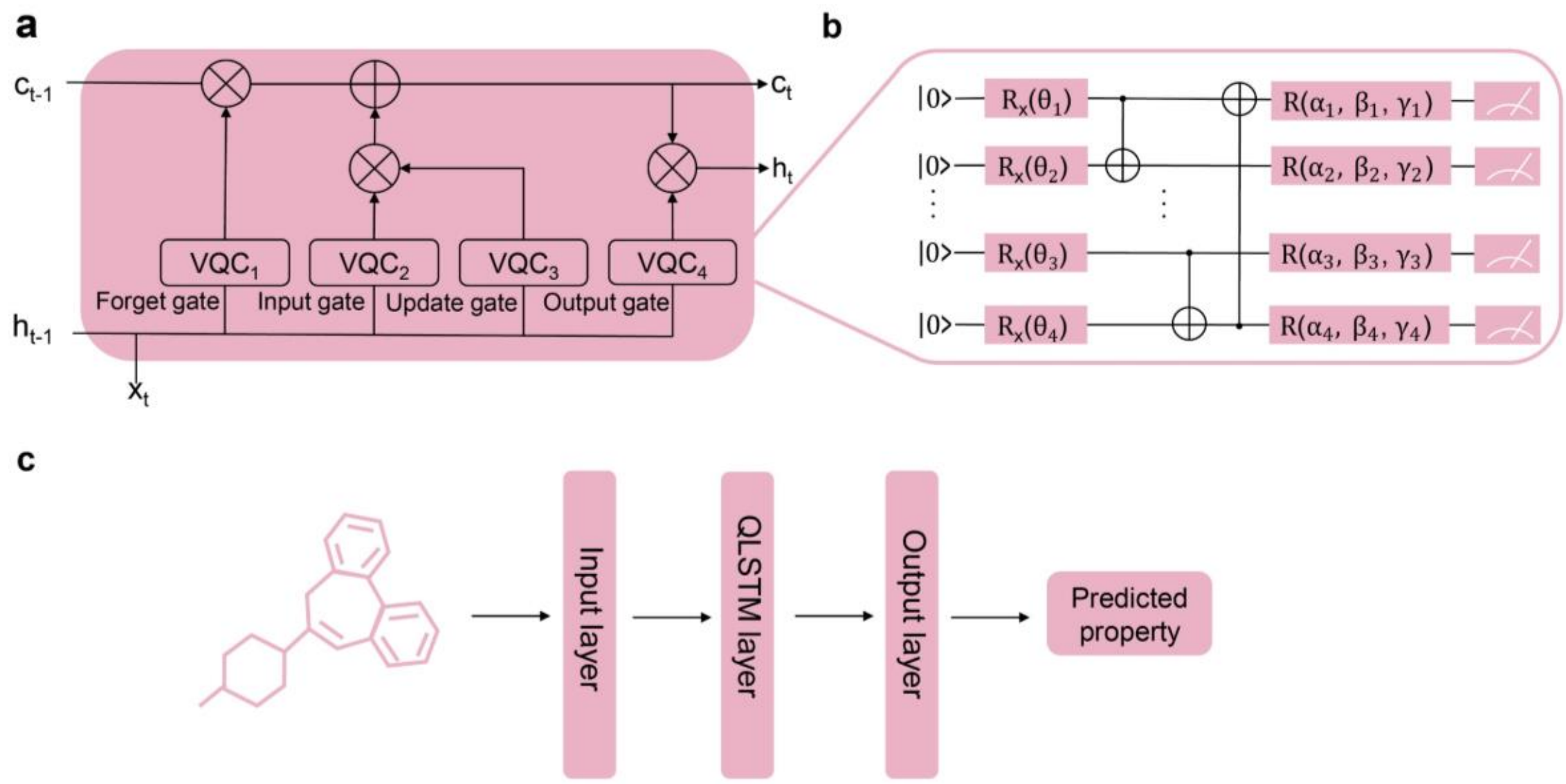

**Figure 1.** The workflow of QLSTM for molecular property prediction. (a) QLSTM model architecture. (b) VQC architecture for QLSTM. The number of qubits and measurements can be dynamically adjusted to meet the specific requirements of the application, ensuring adaptability across different tasks.(c) The whole framework of QLSTM. The input layer transforms the feature dimensions to ensure compatibility with subsequent layers. The QLSTM layer, as the quantum analogue of the classical LSTM, utilizes VQC to process high-dimensional molecular features and capture complex nonlinear relationships. The output layer generates the final prediction, completing the molecular property prediction workflow.

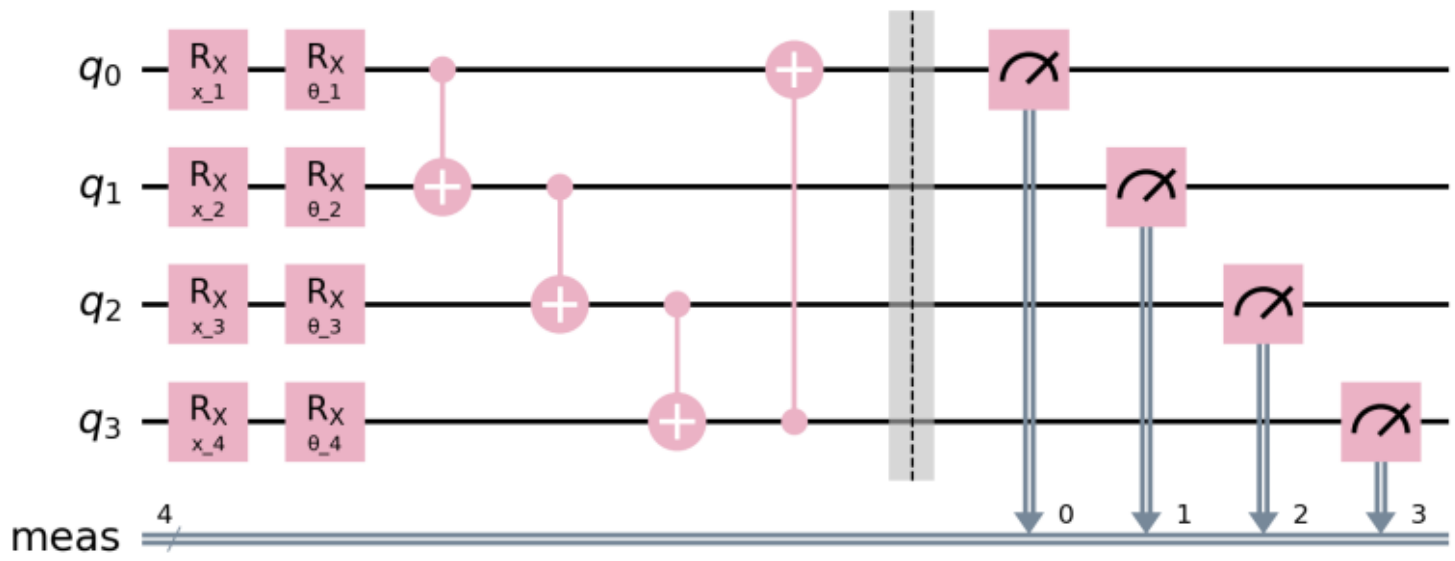

**Figure 2.** Four-qubit quantum circuit. Each qubit is individually adjusted using RX rotation gates and connected via CNOT gates to establish entanglement and interactions between qubits.

The Quantum Neural Network (QNN) is another quantum machine learning model that utilizes PQC to process input data. Each QNN layer consists of rotation gates (RX) and entanglement gates (CNOT), followed by quantum measurements for feature extraction. The Quantum Support Vector Machine (QSVM) utilizes quantum kernel functions based on the ZZFeatureMap to compute similarity between data points in a high-dimensional Hilbert space. To enhance computational efficiency and compatibility with quantum hardware constraints, we applied Principal Component Analysis (PCA) to reduce the input feature space from 1024 dimensions to a lower-dimensional representation matching the number of qubits used in both QNN and QSVM. For a fair comparison, the LSTM model utilized a linear layer to downscale the 1024-bit ECFP molecular fingerprint to match the qubit count, while QLSTM achieved dimensionality reduction through the VQC. All models were trained using three independent data splits with different random seeds, and performance metrics were averaged across these splits to ensure robust evaluation. The RMSprop optimizer[21] was employed, and task-specific hyperparameters for each dataset were manually configured based on empirical observations and prior knowledge. The detailed configurations, including the number of layers, hidden units, learning rates, batch sizes, and dropout rates, are provided in Appendix A (Table S1). The batch size was set to 256, and the training epochs were set to 80.

We also conducted a performance comparison of the QLSTM model under different levels of added noise. To evaluate whether the QLSTM model can be adapted to real NISQ devices, we employed the following score function $s$ to estimate the overall error rate of the QLSTM model[22, 23] on actual quantum hardware. The score function $s$ is defined as:

$$s = 1 - \prod_{j=1}^{d} \left(1 - \left(\frac{\sum_i E_{r_i} N_i}{\sum_i N_i}\right)_j\right)^{m_j} \tag{7}$$

where $N_i$ is the number of quantum logic gate, $E_{r_i}$ is the corresponding error rate of this type of gate, $d$ is the depth of the quantum circuit, the last term, $\left(\frac{\sum_i E_{r_i} N_i}{\sum_i N_i}\right)_j$ is the average error rate in the $j$-th layer, and $m_j$ is the number of gate at circuit layer $j$.

Despite theoretical predictions that accumulated quantum gate and circuit errors may impair model performance, our empirical results reveal that QLSTM consistently achieves high and stable predictive accuracy under varying noise levels. To explain this robustness, we draw upon two key mechanisms from quantum gradient optimization theory, which help account for QLSTM's stability in noisy environments.

Although noise and gate errors typically impact model performance, our experimental results demonstrate that the QLSTM model maintains significant robustness and achieves consistently high predictive accuracy under varying noise levels. This robustness is attributed to two key mechanisms from quantum gradient theory, which help mitigate the impact of noise on gradient updates, thereby ensuring the model's stability and robust optimization in noisy environments.

First, the Parameter-Shift Rule[41], a fundamental method of quantum gradient computation, calculates gradients by shifting quantum circuit parameters to symmetric offsets and measuring the resulting differences in expectation values. This method not only performs well in noisy environments but also improves the stability of gradient calculation by symmetrically eliminating first-order noise effects:

$$\frac{\partial L}{\partial \theta_i} = \frac{1}{2}\left(\langle H \rangle_{\theta_i + \frac{\pi}{2}} - \langle H \rangle_{\theta_i - \frac{\pi}{2}}\right) \tag{8}$$

Where $L$ is the loss function, $\theta_i$ is the model parameter, and $H$ is the Hamiltonian. The symmetric nature of this approach effectively neutralizes first-order noise effects, thereby minimizing their impact on the model training process.

Secondly, the Gradient-Noise Orthogonality Constraint[42] further enhances the robustness of quantum gradient computation. By minimizing the inner product between the gradient and noise, it significantly reduces noise interference in the optimization process. This constraint ensures that even under high noise conditions, the gradient direction remains stable, effectively mitigating the impact of noise on quantum model training:

$$E[\langle \nabla L, \epsilon \rangle] \leq s \cdot T_r(F^{-1}) \tag{9}$$

Here, $E[\langle \nabla L, \epsilon \rangle]$ represents the inner product between the noise and the gradient, and $F$ is the quantum Fisher information matrix that quantifies the orthogonality between noise and gradient.

By introducing quantum gradient theory, we provide a deeper insight into the noise evaluation described in Equation (7). The explanations of the Parameter-Shift Rule and the Gradient-Noise Orthogonality Constraint elucidate how noise[43] affects gradient computation and how these mechanisms counteract its disruptive effects during model training, ensuring robust performance across varying noise levels.

All experiments were conducted on an NVIDIA A100 GPU with 80 GB of memory, running on a 64-bit CentOS v8.5 server equipped with 512 GB of RAM. The QLSTM and QNN models were implemented using Torch Quantum (backed by Qiskit's Aer simulator), and the QSVM was implemented using Qiskit's Aer simulator. Due to hardware constraints, the number of qubits used for QLSTM model comparisons was 2, 4, 8, 12, and 16.

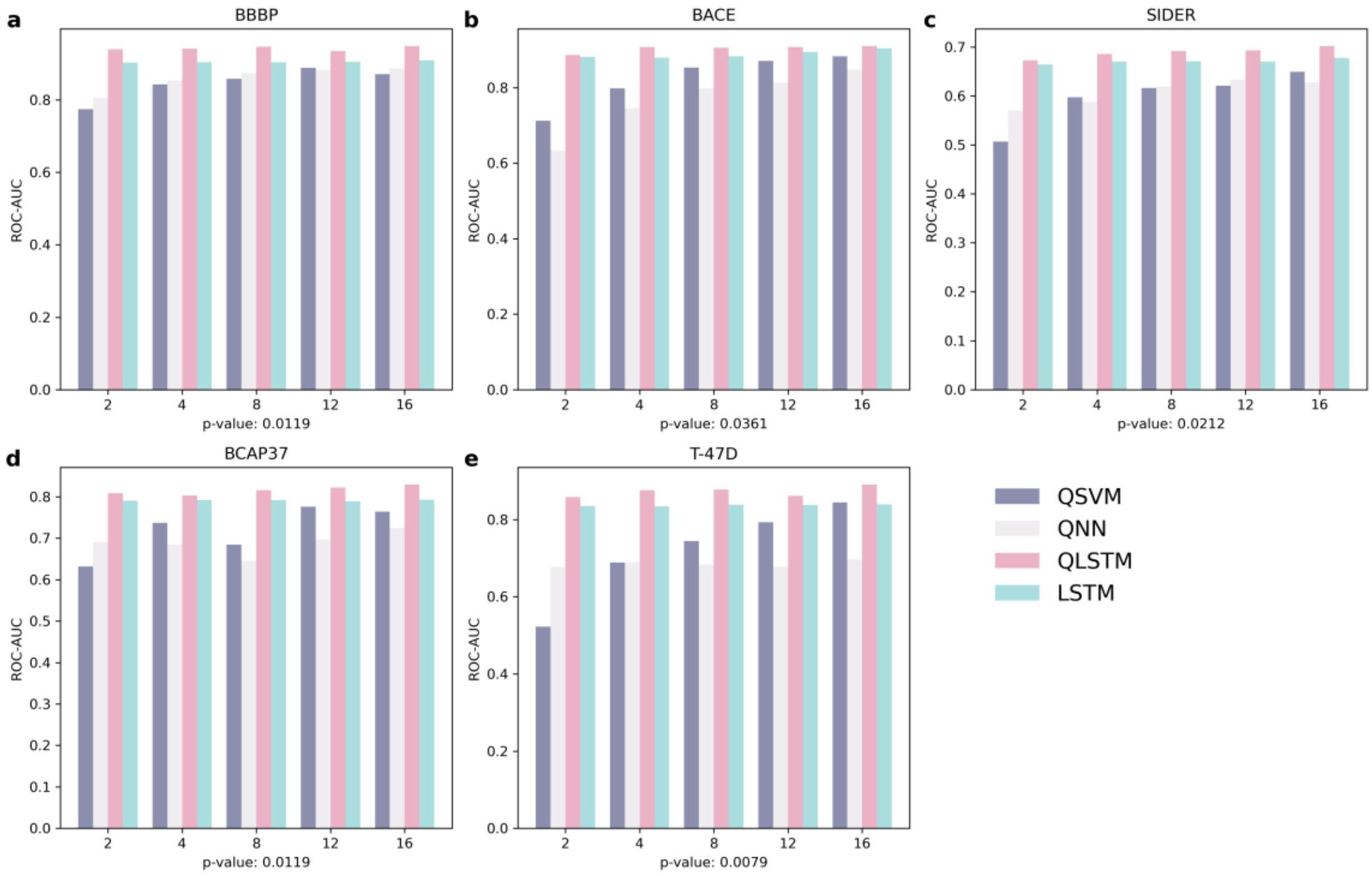

**Figure 3.** ROC-AUC scores of QSVM, QNN, QLSTM, and LSTM on five benchmark datasets: (a) BBBP, (b) BACE, (c) SIDER, (d) BCAP37, and (e) T-47D. Each group of bars shows the models' ROC-AUC scores under increasing qubit counts (2, 4, 8, 12, 16). The p-values from the Mann-Whitney U test, displayed below each subplot, indicate the statistical significance of QLSTM's performance improvements over LSTM. QLSTM demonstrates consistently superior performance across all datasets and qubit settings.

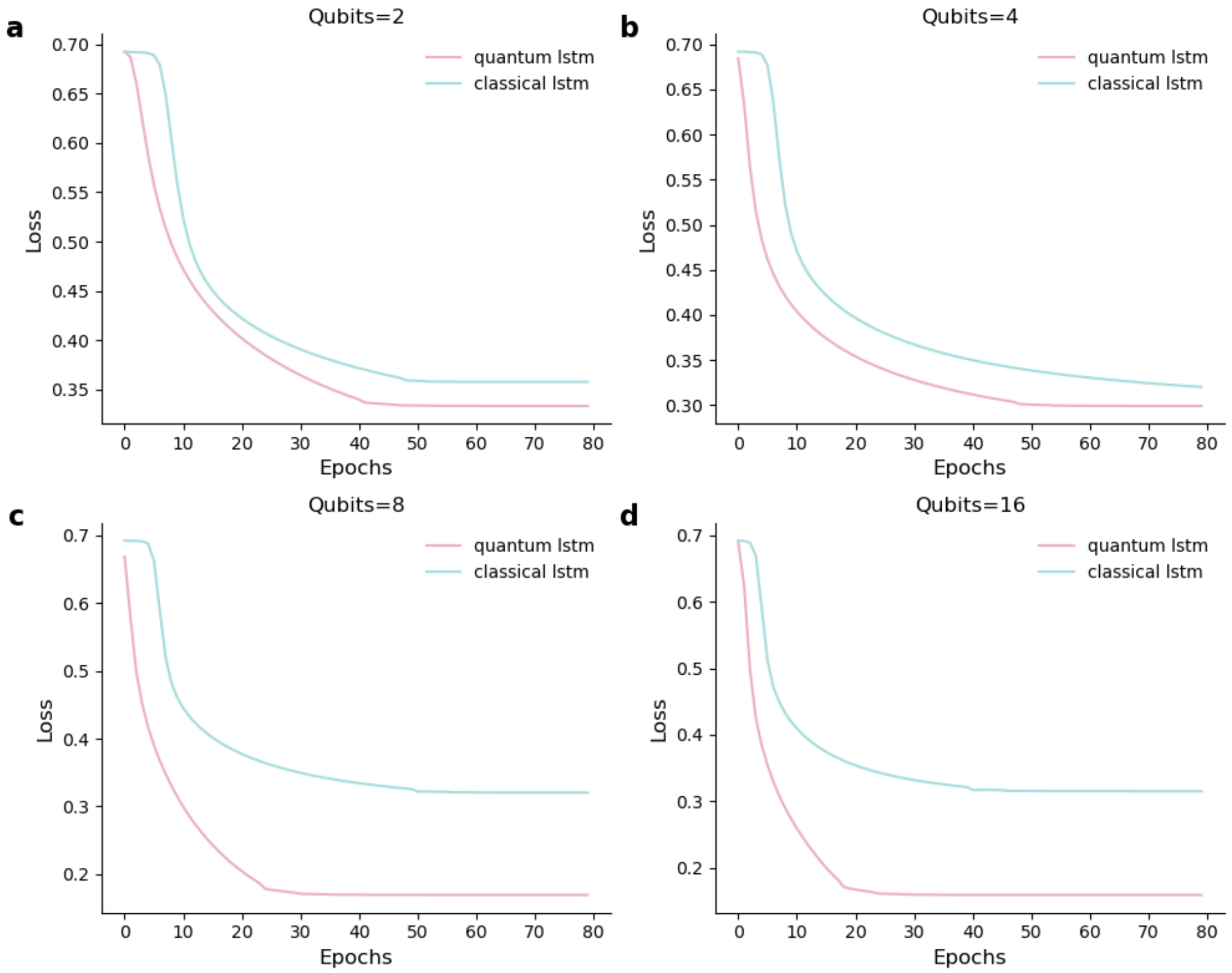

**Figure 4.** Training loss curves of QLSTM and classical LSTM models on the BACE dataset under different qubit settings: (a) 2 qubits, (b) 4 qubits, (c) 8 qubits, and (d) 16 qubits. In all cases, QLSTM demonstrates faster convergence and achieves lower final loss values, highlighting its superior training efficiency.

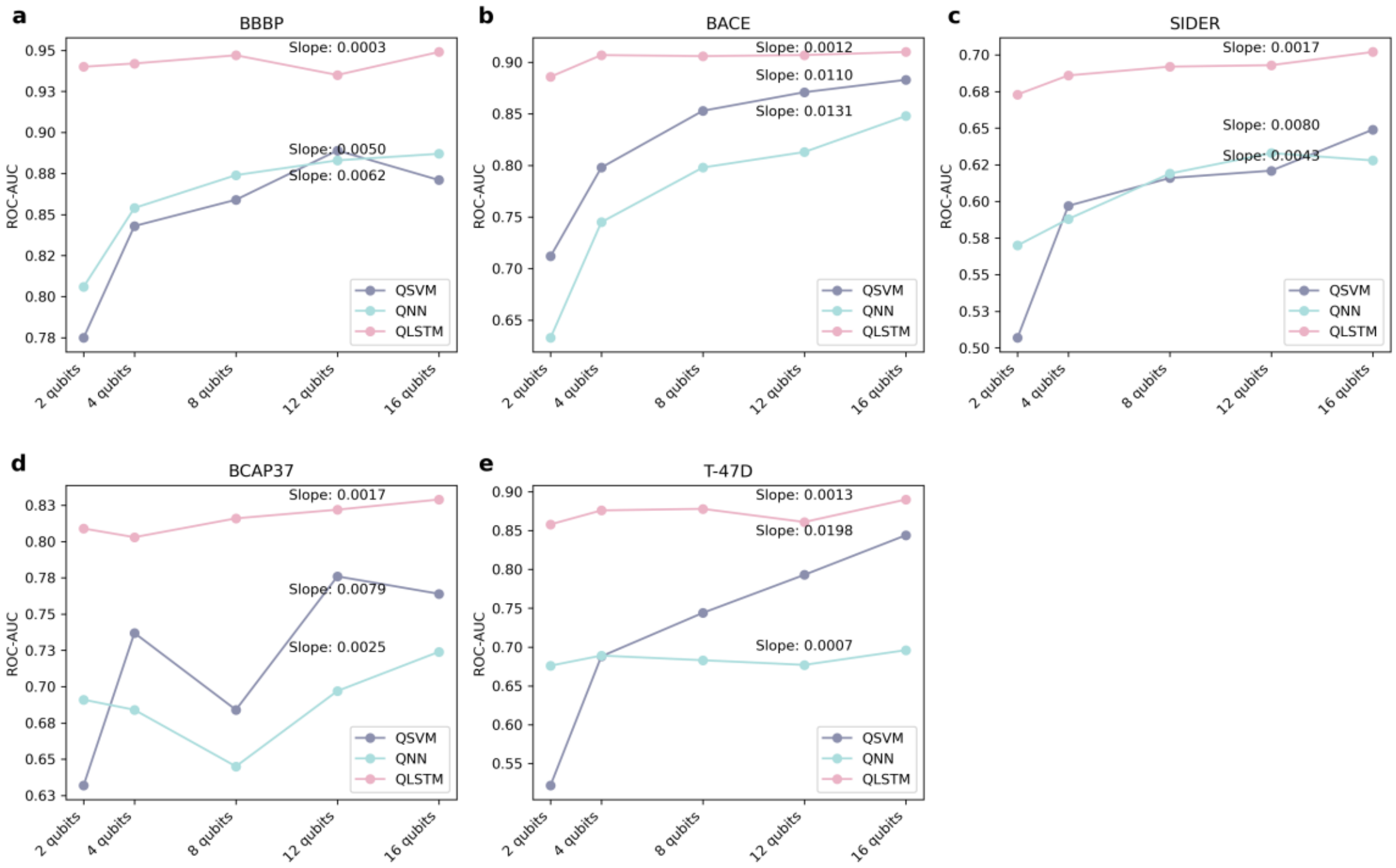

**Figure 5.** ROC-AUC scores of QLSTM, QNN, and QSVM across five benchmark datasets: (a) BBBP, (b) BACE, (c) SIDER, (d) BCAP37, and (e) T-47D. Each line represents the performance trend of a model under different qubit counts (2, 4, 8, 12, 16). The annotated slope values quantify the rate of performance change with respect to qubit scaling. All slopes are positive, suggesting that expanding quantum resources leads to performance gains.

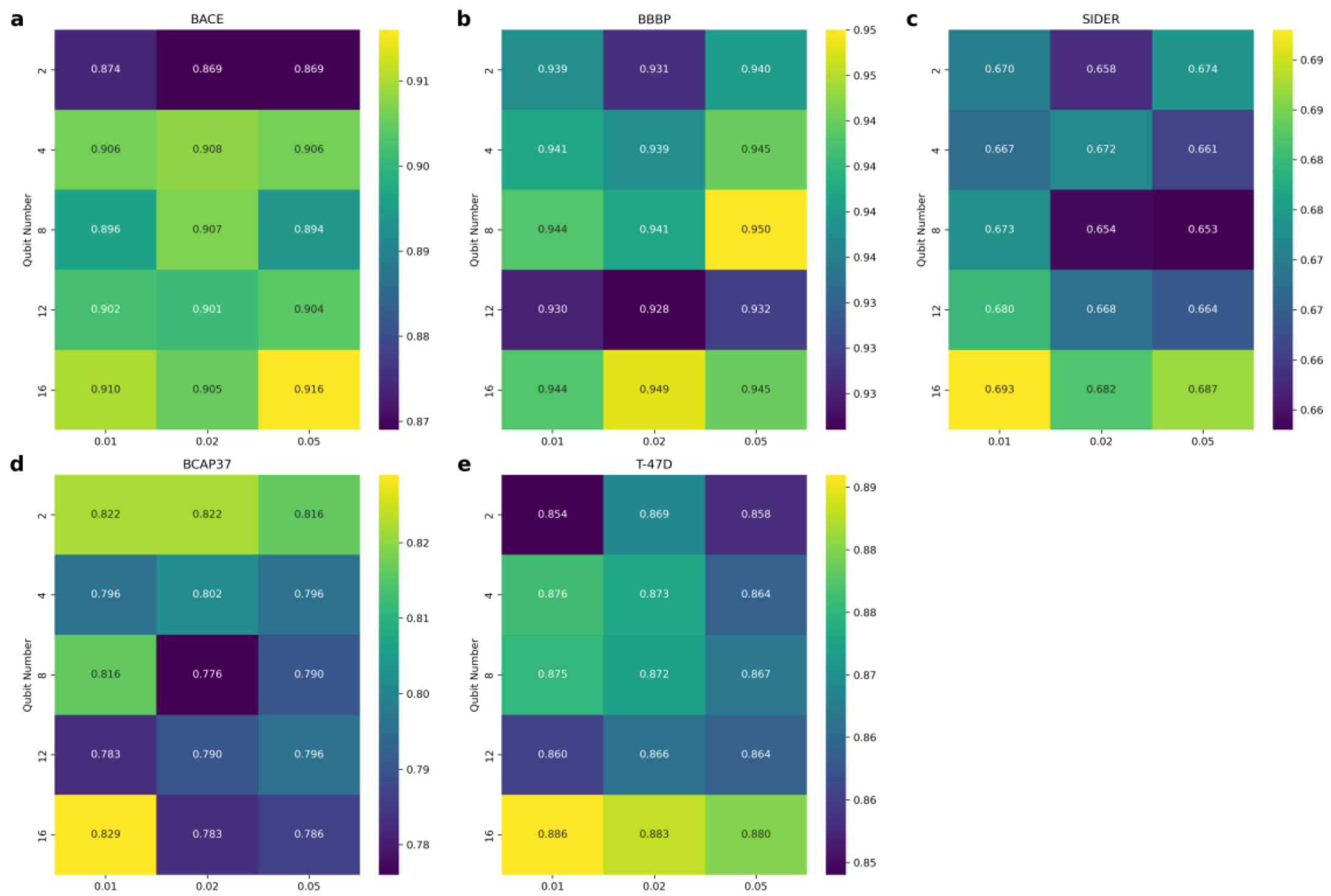

**Figure 6.** Heatmaps showing the ROC-AUC performance of the QLSTM model on five benchmark datasets: (a) BACE, (b) BBBP, (c) SIDER, (d) BCAP37, and (e) T-47D. The experiments were conducted under different noise levels (0.01, 0.02, 0.05) and qubit counts (2, 4, 8, 12, 16). Brighter colors indicate better model performance. The results illustrate that QLSTM maintains robust prediction accuracy even under increased quantum noise.

## Results

To evaluate the impact of qubit count on model performance, we tested QLSTM, QNN, QSVM, and LSTM models with varying qubit counts (2, 4, 8, 12, and 16) across five benchmark datasets (BBBP, BACE, SIDER, BCAP37, and T-47D). These datasets cover a diverse range of drug discovery tasks, including molecular property prediction and anti-cancer activity analysis, ensuring the models' broad applicability across different domains.

The results demonstrate that QLSTM consistently outperforms classical LSTM across all datasets. For high-complexity datasets such as BBBP and T-47D, QLSTM achieved ROC-AUC score improvements ranging from 0.03 to 0.06 over LSTM (Table S2), with statistical significance confirmed by Mann-Whitney U tests ($p < 0.05$) (Figure 3). To evaluate model convergence, we analyzed the loss curves of QLSTM and LSTM. Both models exhibit similar loss reduction trends during the initial training stages. However, QLSTM reached its optimal validation performance in fewer iterations, indicating faster convergence (Figure 4 and Appendix Figures S1-S4). To further assess QLSTM's performance superiority, we compared it with QNN and QSVM models. The results show that QLSTM achieves substantial accuracy improvements across all five datasets, with gains ranging from 9% to 27% over QNN and 10% to 21% over QSVM (Table S3). These findings underscore QLSTM's exceptional capability in handling complex tasks. As shown in Figure S5, QLSTM achieves the highest average score among all models, further emphasizing its overall advantage. Additionally, as shown in Figure 5, the performance of all quantum models (QLSTM, QSVM, and QNN) generally improves as the number of qubits increases, although minor fluctuations are observed in certain cases. This trend suggests a positive correlation between qubit count and model performance. Possible overfitting concerns in the BCAP37 dataset were addressed through 3-fold cross-validation (Table S4), with validation loss stabilizing after 30 epochs (Figure S6), indicating that no significant overfitting occurred. Statistical tests further confirmed the robustness of QLSTM's performance ($p = 0.0159$).

Given the presence of quantum noise in current NISQ devices, we investigated the impact of three levels of bit-flip noise (0.01, 0.02, and 0.05) on QLSTM. The results show that QLSTM exhibits strong robustness, maintaining high performance even under high noise levels (Table S5, Figure 6). For example, on the SIDER dataset with 2 qubits, the ROC-AUC score at a noise level of 0.05 was 0.674, outperforming the score of 0.658 at a noise level of 0.02. Similarly, on the BBBP dataset with 16 qubits, the score at a noise level of 0.05 was 0.945, nearly identical to the score of 0.949 at a noise level of 0.02. This resilience is primarily attributed to the gradient update mechanism, which minimizes noise interference during optimization when the inner product between the gradient direction and noise direction is small[25]. To gain a deeper understanding of the noise mitigation behavior of QLSTM, further exploration and visualization of the gradient update process are needed, which is beyond the scope of this paper and will be explored in future work.

The median error rate for single-qubit quantum gates in IBM quantum processor (Heron) is approximately 0.03%, whereas the median error rate for double-qubit controlled-Z gates is approximately 0.32%[26]. We assess the noise error rate of QLSTM model on real quantum computers, based on the above parameters and Equation 7, resulting in error rates of 0.5%, 1.1%, 2.1%, 3.2%, and 4.4% for circuits with 2, 4, 8, 12, and 16 qubits, respectively. At these noise levels, and even up to 5%, the QLSTM consistently maintains high performance, demonstrating its practicality and potential to drive advances in quantum computing.

## Discussion

Our findings highlight QLSTM's potential in drug discovery, particularly in processing high-dimensional molecular data represented by SMILES strings. Experimental results show that QLSTM significantly outperforms classical LSTM and other quantum machine learning models, including QNN and QSVM. By leveraging parameterized quantum circuits for feature extraction, QLSTM is capable of capturing complex patterns within intricate molecular datasets, making it especially effective for tasks such as molecular property prediction and bioactivity classification. Additionally, QLSTM maintains strong robustness under noisy conditions, possibly due to the stabilizing effect of its gradient update mechanism. This observation warrants further investigation to better understand the underlying principles.

From a training time perspective, the classical LSTM is relatively stable across different input dimensions, with an average training time from 3.16 to 3.36 seconds, while QLSTM shows a significant increase in training time as the number of qubits increases, reaching 680.22 seconds at 16 qubits (Table S6). The main reason for the increase in QLSTM's training time is that the quantum circuits computing in the current QLSTM training is simulated by classical computers, and the simulation time increases exponentially with the number of simulated quantum bits, which leads to an increase in the overall training time[24].

The LSTM's training time is closely related to the input dimensionality *D* and the size of the hidden layer *H*. The computation at each time step primarily involves matrix multiplication from input to hidden layer, with a complexity of $O(D \cdot H + H^2)$[44]. When the input dimensionality *D* is smaller than the hidden layer size *H*, the increase of dimensionality *D* has a relatively small impact on the total computational cost. However, when the input data dimensionality *D* is large, the increase of input dimensionality plays a major role on the total computational overhead[45]. In contrast, the VQC unit in QLSTM training process is quite efficient. Since n qubits can represent a Hilbert space of $2^n$ dimensions, only $\lceil \log_2 \mathrm{D} \rceil$ qubits are required to encode input data of dimensionality *D*. Consequently, as the input data dimensionality *D* increases, the number of qubits grows logarithmically, following *O(log D)*. Additionally, the number of quantum gates and the circuit depth scale linearly with the number of qubits: each VQC consists of 2n RX gates and n CNOT gates, with an overall circuit depth of approximately 2+n, including an input encoding layer, a parameterized rotation layer, and a sequence of entangling operations. Thus, as the input data dimensionality *D* increases, the depth and total number of operations of the quantum circuits increase as

*O(log D)*.

It is also important to consider the efficiency of quantum gate operations on real quantum hardware. For instance, IBM's latest 156-qubit processor (*ibm_fez*) reports an average quantum gate operation time just 68 nanoseconds[46]. For the VQC used in this work, the overall execution is primarily determined by the circuit depth and the individual gate operation time, with some minor contributions from additional overheads such as measurement time. Given this, it is reasonable to infer that the runtime of the VQC scales approximately linearly with the number of qubits[51]. This, coupled with the minimal and reducing gate operation time in quantum hardware, highlight that QLSTM has the potential to substantially reduce computational overhead—particularly in scenarios involving high-dimensional data processing.

To further bolster the practical application of QLSTM in drug discovery, it is essential to consider recent advancements in quantum noise reduction techniques and their potential implications[27, 28]. Quantum error correction (QEC) methods, such as surface codes and concatenated codes, have demonstrated notable success in reducing gate error rates and extending qubit coherence times[29, 30]. In parallel, error mitigation techniques like zero-noise extrapolation (ZNE) and probabilistic error cancellation (PEC) have enabled more accurate computations on near-term noisy quantum devices[31]. A significant milestone in this trajectory is the development of Google's Willow chip, which has achieved substantial reductions in logical qubit error rates, marking a key advance in fault-tolerant quantum computing[52]. Moreover, photonic quantum computing platforms[39, 40] have made considerable progress in enabling high-speed gate operations, offering additional pathways for scalable quantum architectures. As these technologies continue to evolve, the overall error rates in quantum systems are expected to decline, paving the way for more precise and efficient quantum computations in real-world applications.

Although QLSTM has not yet been implemented on real quantum hardware, its strong performance in molecular property prediction and bioactivity classification highlights its potential for drug discovery applications. Its modular design and compatibility with high-dimensional molecular data make it a promising candidate for early-stage screening tasks, where it could be integrated into existing drug development pipelines to assist in identifying candidate compounds. The scalable nature of its circuit design suggests that QLSTM stands to benefit from quantum acceleration as hardware capabilities advance—especially for small- to medium-scale, complex datasets. With continued progress in quantum hardware development and error mitigation strategies, QLSTM is well positioned to transition from simulation to real-world implementation, enabling broader adoption of quantum machine learning in the life sciences.

## Conclusion

In summary, we have successfully implemented the QLSTM model in drug discovery, achieving superior performance and faster convergence compared to the classical LSTM. Notably, as the number of qubits increases, the accuracy of QLSTM improves. Furthermore, our analysis of noise resilience reveals that QLSTM maintains robust performance across varying noise levels, suggesting its feasibility on current NISQ devices. Given the estimated noise error rates of existing quantum hardware, our findings highlight the practicality of executing QLSTM on real quantum processors. This work marks a significant step forward in quantum machine learning for scientific research, paving the way for future advancements in hybrid quantum-classical models.

**Acknowledgements**

Central government guide local science and technology development funds No: 2023JH6/100100065. This work was partly supported by Tianyan Quantum Computing Cloud Platform.

**Conflict of Interest**

The authors declare no conflict of interest.

**Code Availability Statement**

The code that supports the findings of this study are available from the corresponding author upon reasonable request.